# Coulomb barrier creation by means of electronic field emission in nanolayer capacitors

Eduard Ilin, Irina Burkova, Timothy Draher, Eugene V. Colla, Alfred Hübler (deceased), and Alexey Bezryadin
Department of Physics, University of Illinois at Urbana-Champaign, Urbana, IL, 61801, USA

**Abstract:** *The main mechanism of energy loss in capacitors with nanoscale dielectric films is leakage currents. Using the example of Al-Al$_2$O$_3$-Al, we show that there are two main contributions, namely the cold field emission effect and the hopping conductivity through the dielectric. Our main finding is that an application of a high electric field, ~0.6-0.7 GV/m, causes electrons to penetrate the dielectric. If the temperature is sufficiently low, such electrons become permanently trapped in the dielectric. To achieve a strong charging of the dielectric, the voltage needs to be high enough, so that a field emission occurs from the cathode into the dielectric. Such a strongly charged dielectric layer generates a Coulomb barrier and leads to a suppression of the leakage current. Thus, after the dielectric nanolayer of the capacitor is charged, the field emission and the hopping conductivity are both suppressed, and the hysteresis of the I-V curve disappears. The phenomenon is observed at temperatures up to ~225 K. It would be advantageous to identify insulators in which the phenomenon of the Coulomb barriers persists even up to the room temperature, but at this time it is not known whether such dielectrics exist and/or can be designed.*

## I. INTRODUCTION

Metal-insulator-metal capacitors with nanometer-scale dielectric layers have attracted a lot of attention recently, due to their various possible technological applications [1,2,3,4,5]. The central element of such capacitors is the nanometers-thick insulating film that defines the ability of such systems to withstand strong electric fields without allowing leakages. Previous studies have shown that the leakage current and the breakdown voltage depend on the choice of the materials for the capacitor plates [6,7]. Such nanoscale dielectrics are important in many spheres of technology, notably in electronic energy storage systems, such as metal-dielectric-metal capacitors [5,8,9,10,11,12]. Nanoscale dielectrics play a key role in the computer technology, which relies heavily on the performance of ultra-thin gate oxides in the field effect transistors [13,14,15,16]. Yet, all known insulating barriers exhibit some level of conductivity, which is generically called "leakage". This leakage occurs even at relatively low electric fields of less than 2 MV/cm [1,17,18,19,20,21], so low that the tunneling effect and the field emission effect are negligible. This leakage conductivity causes significant energy losses in nanolayer capacitors, as well as the Joule heating in field effect transistors. Thus, it is highly desirable to achieve a better understanding of the processes responsible for the electrical conductivity of dielectric films and to develop new methods of suppression such leakages.

Here we report on a series of measurements illustrating that the low-voltage leakage in nanolayer capacitors can be suppressed by introducing bulk charges into the dielectric film. The "low voltage" term is understood as such low voltage and such low electric field that the field emission current is negligible. In the low voltage regime, the conductivity is due to the electron diffusion or hopping. On the contrary, the charge injection in the dielectric requires a high voltage bias, so high that the field emission current becomes significant. As electrons pass through the capacitor dielectric, some of them get permanently localized on defects and the dielectric traps. Such permanent localization or trapping of the charges takes place only if the temperature is sufficiently low, and the thermal energy is not enough to push the electrons out of the traps (e.g., at 77 K). Such trapped charges reduce the leakage current. The reduction of the leakage is explained by the Coulomb repulsion effect between the electrons trapped in the dielectric film and the mobile electrons participating in the leakage current. In this respect, our results provide a different view on the bulk charges (space charges). Previously it was observed that bulk charges in the dielectric can increase the leakage current through the dielectric, in a form of space charge limited currents [22]. Our focus is different. We report that a reduction of the leakage current can also occur under certain conditions, namely when the injected space charges are trapped and present a Coulomb barrier to the electronic flow.

Previously it was observed that electrons can penetrate the dielectric and thus can increase the charge storage ability of capacitors with a few-nanometer-thick dielectric films [23,24,25,26]. But it was concluded that the charges are not permanently present in the dielectric since the temperature was too high for permanent trapping of the electrons. Namely, at room temperature, the electrons leave the insulating layer of the capacitor as the voltage is reduced. Here we report that such charging of the dielectric can be permanent if the temperature remains low, namely less than about ~200 K for the type of dielectric tested. At such low temperatures, the low-voltage leakage becomes undetectable and the threshold for the beginning of the field emission current increases. Such finding is explained by a Coulomb barrier, as stated above.

## II. EXPERIMENTAL DETAILS

We fabricate our metal-insulator-metal (Al/Al$_2$O$_3$/Al) capacitors on glass substrates. The plates of the capacitor are fabricated out of 99.99% pure Al by thermal evaporation in a vacuum of ~10$^{-5}$ Torr. The top and bottom plate of the capacitor were 25 nm thick in most devices. After the initial deposition of the bottom Al plate, the samples were transferred into an atomic layer deposition (ALD) system (Savannah S100 by Cambridge NanoTech Inc.), where the whole surface was homogeneously covered with alumina, using trimethylaluminum (TMA) and water vapor precursors at 80°C. The thicknesses of ALD grown oxide, $t_{oxide}$, were 7, 10, 15, 20, 25, 30, 50 and 100 nm in the tested samples. The capacitor area was $A=1$ mm$^2$. The time of exposure of the bottom Al film in the atmospheric oxygen during the sample transfer into the ALD chamber did not exceed 1 hour, thus the thicknesses of naturally grown alumina were ~3 nm [27,28].

A chip carrier with the sample was mounted in the $^4$He vacuum-sealed stainless-steel dipstick. The dipstick was vacuum pumped down to ~10$^{-4}$ Torr and then filled with He gas at a pressure of ~10$^{-3}$ Torr needed for thermal equilibration in cryogenic experiments. To shield the samples from external electromagnetic noise, the samples were placed into a Faraday cage located inside the dipstick. Such electromagnetic shielding is essential to obtain high accuracy measurements of the current in the pA or sub-pA range. Low temperature measurements were performed as the dipstick was submerged either into a liquid $^4$He cryostat or into a liquid nitrogen bath.

The measurements were controlled and recorded using LabView software. The "high" voltage terminal was always connected with the top plate of the capacitors, which means that the positive voltage corresponds to the positive potential on the top plate. Note that the top and bottom plates are nominally identical, however, the local roughness could be higher in the top plate, since it is located over the bottom Al and the alumina dielectric, which are morphologically rough since they are not crystalline materials (insert, Fig.1b). To obtain the current-voltage dependence ("I-V curve"), the voltage was applied in small discrete steps of ~0.1 V. The time delay between the voltage steps was 300 seconds for the capacitors with 20, 25, 30, 50 and 100 nm alumina thicknesses and 100 seconds for the 7, 10 and 15 nm alumina thickness. The time delay is needed to detect the true leakage current. Generally speaking, the current through the capacitor includes three components[26], namely, (1) the charging current of the plates, (2) the dielectric charging current, and (3) the leakage current. The first one decreases exponentially with time; the time constant is determined by the characteristics of the circuit, namely the capacitance, $C$, and the standard resistor, $R_{st}$, inserted in series with the capacitor. The capacitance of our samples was between 1 nF and 9 nF, for the dielectric layer between 100 to 7 nm thick, respectively. The series resistor was $R_{st}$ = 10.6 MΩ, thus the plates charging time was $\tau = R_{st}C \approx 0.01 - 0.09$ s. Since the charging current drops exponentially, its contribution to the total current becomes negligible after a few seconds. On the contrary, the dielectric-charging current continues for hundreds of seconds[26]. Thus, to detect the true leakage current, long equilibration times were necessary between the voltage application and the subsequent measurement of the response current (listed above). With these delays we were able to ensure that the measured current represents the true leakage and not some sort of transient process. The voltage on the capacitor (the "sample") was calculated using the formula $V_S = U - IR_{st}$, where $R_{st}$ is the series resistor, $I$ is the measured current in the circuit and $U$ is the applied voltage. This formula accounts for the fact that the voltage $U$ is not applied directly to the capacitor, but to the capacitor and the resistor $R_{st}$ connected in series.

The measurements of the current in the circuit were performed in most cases using Keithley 6517B. This device is equipped with an adjustable voltage source, which also provides a controlled voltage biasing. Alternatively, we have used a National Instruments data acquisition board (DAB) NI-USB 6216. The circuit corresponding to such DAB measurements is shown on Fig.1a. The switch has to be in position "1" in order to charge the capacitor. The discharge process takes place when the switch is turned to position "2". Both charging and discharging are carried through a series standard resistor of $R_{st1}$=1 GΩ. The current was calculated by Ohm's law formula ($I=V_R/R_{st1}$). The bias voltage $U$ is also provided by the same DAB, namely NI-USB 6216.

## III. RESULTS

Room temperature measurements are shown in Fig.1a. In these experiments, the current flowing into the capacitor was measured, as the voltage was increasing with a step size $\Delta U = 0.1$ V and time delay of 100 or 300 s. Since the voltage was increasing very slowly, the presented current represents the leakage through the dielectric layer. The main feature of all the I-V curves is that initially the current increases slowly, but, as the voltage exceeds some sample specific threshold, the current exhibits a very sharp upturn. The threshold voltage, at which the strong increase of the leakage current begins, is denoted $V_{th}$. For example, the sample with the oxide thickness of 100 nm (curve #8) has a threshold of $V_{th}$=30 V. Obviously, for the applications where the leakage is supposed to be low, the bias voltage should be lower than the high leakage threshold $V_{th}$.

A universal behavior is observed as we plot the current versus the electric field, Fig.1b. In this figure one can identify when the strong leakage current regime begins, namely at $E_{th}$~0.33 GV/m. This electric field sets the limit for the application, in which a low leakage is required.

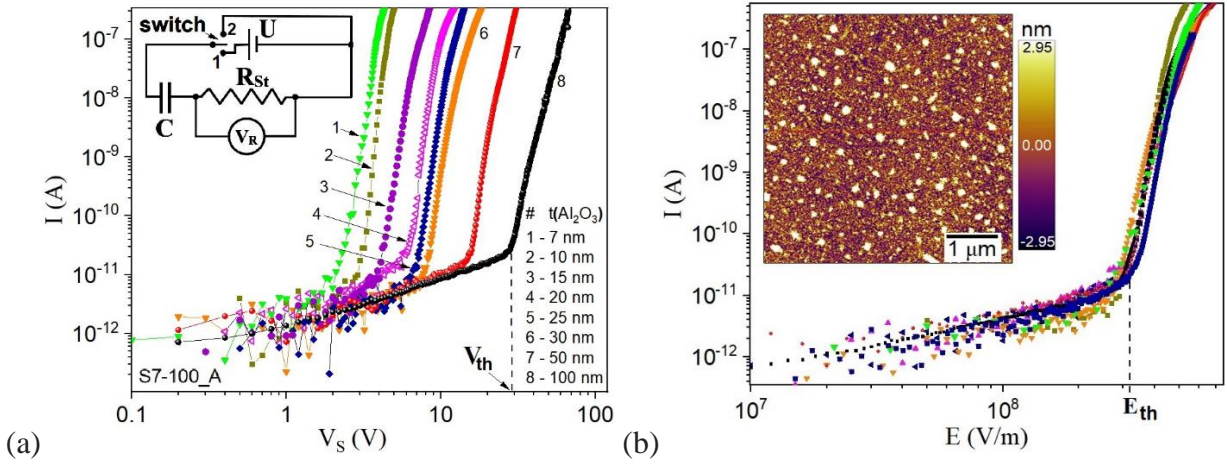

Figure 1. (a) I-V plots for the leakage current in Al/Al$_2$O$_3$/Al capacitors with different thicknesses of the dielectric, measured at 295 K. The dielectric layer thickness for the eight tested samples is indicated on the graph. The threshold voltage, $V_{th}$, separates the low bias regime and the high bias regime. At high bias, $V>V_{th}$, the field emission effect is dominant and much stronger than the leakage detected at low bias, $V<V_{th}$. (a, insert) Electronic scheme of the setup: as the capacitor is charged from the voltage source U, the current, I, is monitored by measuring $V_R$ and computing the current using Ohm's law $I=V_R/R$. The voltage $V_R$ was measured using National Instruments data acquisition board NI-USB 6216. (b) All the curves of graph (a) are plotted versus the electric field. A universal behavior is observed, where the current density is defined by the strength of the electric field. The area of all reported capacitors was 1 mm$^2$. The threshold electric field is defined as $E_{th}=V_{th}/t_{oxide}$, where $t_{oxide}$ is the dielectric thickness. (b, insert) Atomic force microscope (AFM) image of the bottom plate of a capacitor.

To detect the field emission current, we plot the logarithm of the current density normalized by the second power of the electric field versus the inverse electric field (Fig.2a). The formula for the cold field emission tunneling current density is[29]:
$J = e^3E^2/(8\pi h e \phi_b) \cdot \exp[-8\pi(2em^*)^{1/2}\phi_b^{3/2}/3hE]$, where $J$ is the current density, $e$ is the electronic charge, $h$ is the Planck's constant, $\phi_b$ is metal-insulator energy barrier height, $m^*$ is effective electron mass in the insulator. The current density $J$ is calculated as $J = I/A$, where $A$ is the area of the capacitor and $I$ is the measured current. The signature of the field emission effect is that the I-V curve appears linear if it is plotted in the coordinates $y=\ln(J/E^2)$ vs $x=1/E$. The linear behavior seen in Fig.2a gives strong evidence that the leakage at higher voltages is due to the field emission effect. The straight line is a fit plotted according to the expression given above.

The slope of the linear part of the $y$ vs $x$ curve (Fig.2a) is given by the formula $dy/dx=-8\pi(2em^*)^{1/2}\phi_b^{3/2}/3h=-6.83\cdot 10^9(m^*/m_0)^{1/2}\phi_b^{3/2}$. The value of the slope is defined by the best fit. Then, the effective barrier was determined assuming that the effective mass is[30] $m^*=0.05m_0$, the result being $\phi_b=3.25$ eV.

At higher values of *x* (lower values of *V* and *E*), a minimum in the y-x plot is observed at $E_{th}=0.33$ GV/m, which indicates that the physical mechanism determining the electrical current in the insulator changes. Previously, such a minimum has been linked to a transition from the field emission (at higher voltages) to direct tunneling (at lower voltages) [31]. Yet, the observed weak (weaker than exponential) dependence of the leakage current on the dielectric thickness (Fig.1b) strongly suggests that the direct tunneling is not essential in our sample. Note also that all curves collapse on one curve (Fig.1b). Thus, the mechanism should be the same for all of them. Yet, tunneling of electrons over a distance of 100 nm in a dielectric is very unlikely, so the direct tunneling is negligible in this series of samples. Our conclusion that the direct tunneling is negligible in our samples is in agreement with previously published results (Ref. 19) where it was shown that the direct tunneling is significant only through the alumina films which are thinner than approximately 3 nm. Our thinnest films were 7 nm.

We have tested different mechanism, including electron diffusion, direct tunneling, Schottky emission, thermionic-field emission, Poole-Frenkel emission [17,29]. It appears that the most suitable explanation for the leakage current (at low electric fields, $E<0.33$ GV/m), which does not lead to unphysical assumptions about the fitting parameters, is the hopping conductivity. The theory of the hopping conductivity is outlined in Ref.29. The hopping conductivity formula is $J=eanv \cdot \exp[(eaE-E_a)/kT]$, where *a* is the mean hopping distance, *n* is the electronic density, *v* is the frequency of thermal vibration of electrons at trap sites, *T* is the absolute temperature, *k* is Boltzmann's constant, and $E_a$ is the activation energy. Based on the fact that trapped electrons get released at ~225 K (see below), we suggest that the activation energy is of the same order of magnitude.

As it is shown in Fig.2b, the log(*J*) versus *E* plot is linear in the range $\sim 0.5 \cdot 10^8$ V/m$<E<3.3 \cdot 10^8$ V/m, which is in agreement with the hopping conductivity formula given above. From the slope of the linear fit we determine the average hopping distance in $Al_2O_3$ as $a \approx 1.6$ nm. If one assumes a homogeneous distribution of the traps in the dielectric, then one can estimate the concentration of the electronic traps as $n_t=1/a^3 \approx 2 \cdot 10^{26}$ m$^{-3}$. This is comparable with the estimate of the trap concentration being in the range $10^{25}$-$10^{26}$ m$^{-3}$ given in Ref. 32. The atomic density of the amorphous ALD-deposited alumina[33] is $\sim 10^{29}$ m$^{-3}$. It follows that there is roughly one trap per one thousand atoms.

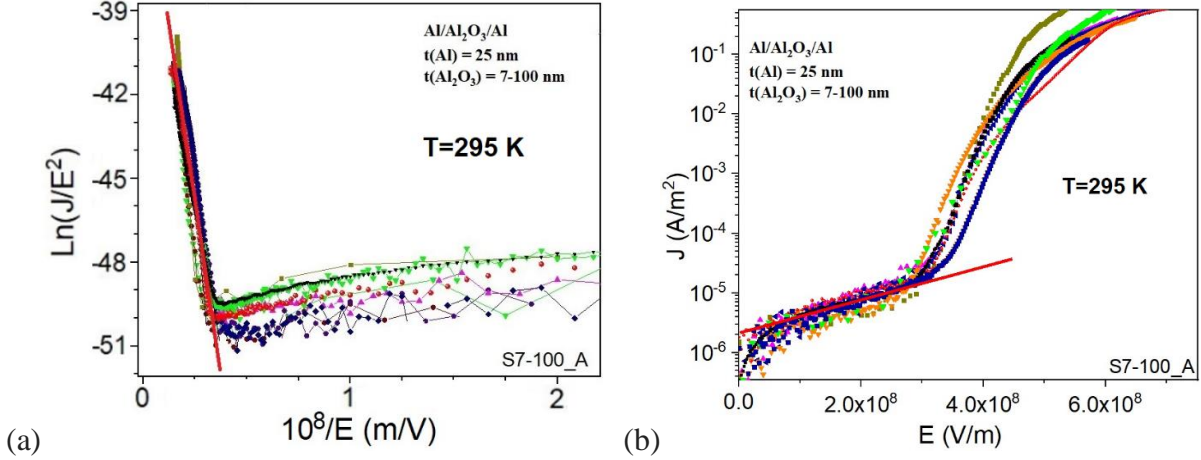

*Figure 2. Experimental results (connected symbols) for the leakage current versus voltage measurements on the eight capacitors with different dielectric layer thickness, varying in the range between 7 nm and 100 nm. (a) These are FN field emission plots in the coordinates $y=\ln(J/E^2)$ versus $x=1/E$, where J is the current density in the dielectric layer and E is the electric field. (b) These are plots made to reveal the hopping conductivity in the coordinates $\log(J)$ versus E. See text for details.*

We have also compared the leakage currents occurring at various temperatures. The results are shown in Fig.3. At low voltages ($V<V_{th}$ or $E<0.33$ GV/m) the leakage current is low and only visible at room temperature, while at cryogenic temperatures this sub-threshold current appears completely suppressed. The threshold voltage, $V_{th}$, is the voltage at which a sharp upturn of the leakage occurs, due to the onset of the field emission current, i.e., tunneling from the cathode into the dielectric layer. This threshold voltage increases significantly with cooling from $V_{th}$~3 V at room temperature up to $V_{th}$~6 V at the liquid helium temperature (4.2 K). We propose that such increase takes place because of the charging of the dielectric layer with external electrons arriving from the negatively charged plate of the capacitor. These injected electrons become trapped on defects of the insulating layer [26,34] (Fig.3a) and contribute to the energy barrier experienced by the tunneling electrons. Such extra barrier, termed Coulomb barrier, is due to the Coulomb repulsion between the electrons trapped on the defect in the dielectric and the electrons participating in the tunneling field emission current.

Key results are obtained as the capacitors were repeatedly charged and discharged (about 7 times at each temperature). We will first discuss the room temperature measurements. At room temperature, a hysteresis is present in the V-I curves (Fig.3). The origin of the hysteresis is that the electrons penetrate the dielectric as the voltage is increased and leave the dielectric as the voltage is reduced to zero[26]. Therefore, the measured total current is elevated as the voltage is ramped up and the current flowing into the capacitor is reduced as the voltage is lowered, due to the backflow of the electrons escaping from the dielectric layer, as was previously reported in Refs. [26,34].

The cycling of the applied voltage, performed at room temperature, does not make a significant change in the behavior, presumably because the external electrons can enter the dielectric as the voltage is ramped up and then they can leave the dielectric as the voltage is ramped down, since thermal fluctuations help them to escape from the electronic traps in the dielectric film. This is the case for the curve #1 in Fig.3.

To summarize, at room temperature, the V-I curves are hysteretic but reproducible, i.e., the V-I obtained in the first voltage sweep cycle is virtually identical to the V-I curve measured in the second and subsequent cycles.

A qualitatively different, and in some sense opposite, behavior is observed at cryogenic temperatures: The V-I curves are hysteretic only in the first voltage sweep cycle and loose hysteresis completely in the subsequent cycles.

To illustrate this, we will now discuss in detail the behavior observed at $T=77$ K. At this cryogenic temperate, the hysteresis is only seen in the first cycle and then disappears in the subsequent cycles (curve #2 in Fig.3). The threshold also changes, namely it is $V_{th1}(77K)\sim 4$ V in the first charging cycle and increases to $V_{th2}(77K)\sim 5.1$ V in the subsequent cycles. Thus, the ratio is $V_{th2}(77K)/V_{th1}(77K)=1.27$. The resulting effect is that at high voltage ($V>V_{th}$) the leakage current is lower in the second and subsequent charging cycles. After the second cycle, the V-I curve does not change significantly with subsequent charging-discharging cycles.

A similar behavior is observed when the sample temperature is further reduced to 4.2 K (see curve #3 in Fig.3). During the first charging cycle, a hysteresis is visible. In the second and subsequent cycles, the V-I curves are reproducible and exhibit no hysteresis. The threshold in the first charging cycle is $V_{th1}(4.2K)\sim 5.2$ V, and it increases to $V_{th2}(4.2K)\sim 6$ V in the subsequent cycles. Thus, the ratio is $V_{th2}(4.2K)/V_{th1}(4.2K)=1.15$.

We propose that the reason for the disappearance of the hysteresis and the decrease of the leakage current observed after the first charging-discharging cycle at $T=77$ K and $T=4.2$ K is that most of the electronic traps are filled within the first charging cycle and, as the voltage is reduced, the electrons remain trapped on the defects in the dielectric. Thus, in the second and subsequent cycles, the electrons trapped in the dielectric present a Coulomb barrier for the leakage current. Moreover, since at low temperatures the electrons which have been trapped remain trapped, no hysteresis occurs on the V-I curve.

After a repetition of a few (6-8) charging-discharging cycles at 4.2 K, we warmed the sample up to 77 K. To our surprise, the sample "remembers" that it was conditioned at 4.2 K and exhibits the same exact V-I curve as was observed at 4.2 K. Thus, we conclude that the thermal energy at $T=77$ K is not large enough to shift the electrons which filled the traps during the previous, lower temperature, charging cycle at $T=4.2$ K. To verify this conclusion, compare the curve #4, taken at 77 K, and the curve #3, which was taken at 4.2 K. These two curves coincide, except in the first cycle (at 4.2 K) in which the traps are being filled. Thus, it appears that the charge traps filled at a much lower temperature remain filled as the sample warms up to $T=77$ K. Apparently, the thermal energy at $T=77$ K is lower than the characteristics trapping energy.

It is interesting to note that by cooling to $T=4.2$ K, we were able to introduce additional charges in the dielectric and thus further reduce the leakage, as well as increase the threshold. These changes remain present even after the sample was warmed up back to $T=77$ K. We speculate that at lower temperatures leakages of all types are further reduced, so that it becomes possible to create higher electric fields, which strengthen the charge injection in the dielectric, without causing too much leakage and associated Joule heating.

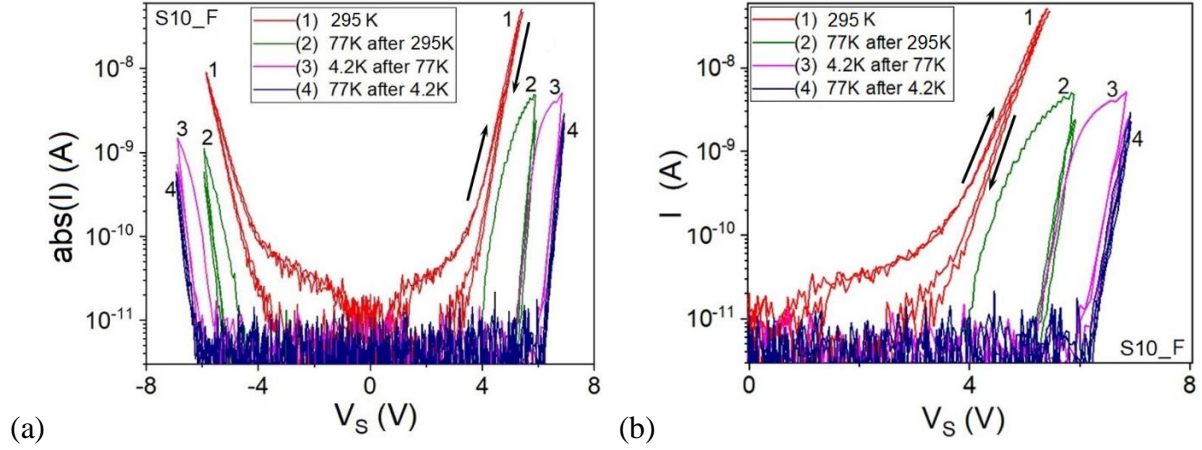

*Figure 3. Leakage current vs. sample voltage $V_s$ for the Al/Al$_2$O$_3$/Al capacitor with the 10 nm Al$_2$O$_3$ thickness at different temperatures. (a) and (b) show the results of the same series of measurements, but in (b) we zoom in only on the positive bias to see the details better. Four measurements are shown, namely the room temperature curve (1), the liquid nitrogen temperature curve (2), and the liquid helium temperature curve (3), and then one more liquid nitrogen temperature measurement (4).*

So far, we have discussed the effect of the dielectric charging using an example of a 10 nm thick dielectric layer capacitor. In what follows we show that the dielectric charging occurs analogously in capacitors with various thickness of the dielectric. The results are shown in Fig.4, in which the left three graphs (a, b, and c) represent the 77 K tests; the right three graphs (d, e, and f) represent the 4.2 K tests. The top graphs (Fig.4(a) and Fig.4(d)) show how the threshold changes when the samples are cooled, for the dielectric thickness ranging between 10 nm and 100 nm. The effect is slightly stronger for thinner dielectric layers. The second row of graphs (Fig.4(b) and Fig.4(e)) represents the ratio of the thresholds corresponding to the charged dielectric layers and not charged dielectric layers, at a given temperature. This ratio is typically near ~1.4, and it becomes somewhat larger for thicker samples. Finally, the bottom two graphs (Fig.4(c) and Fig.4(f)) show the ratio of the charged insulator threshold, measured at the corresponding cryogenic temperature, and the room temperature threshold. The ratio is about ~1.65 for 77 K and ~1.9 for 4.2 K. The ratio of the threshold voltages, $V_{th2}(4.2K)/V_{th}(295K)$~1.9, is roughly independent of the dielectric film thickness, for the thickness range tested. The same is true for the ratio $V_{th2}(77K)/V_{th}(295K)$~1.65. This thickness independence is visualized by the horizontal fit lines in Fig.4(c) and Fig.4(f).

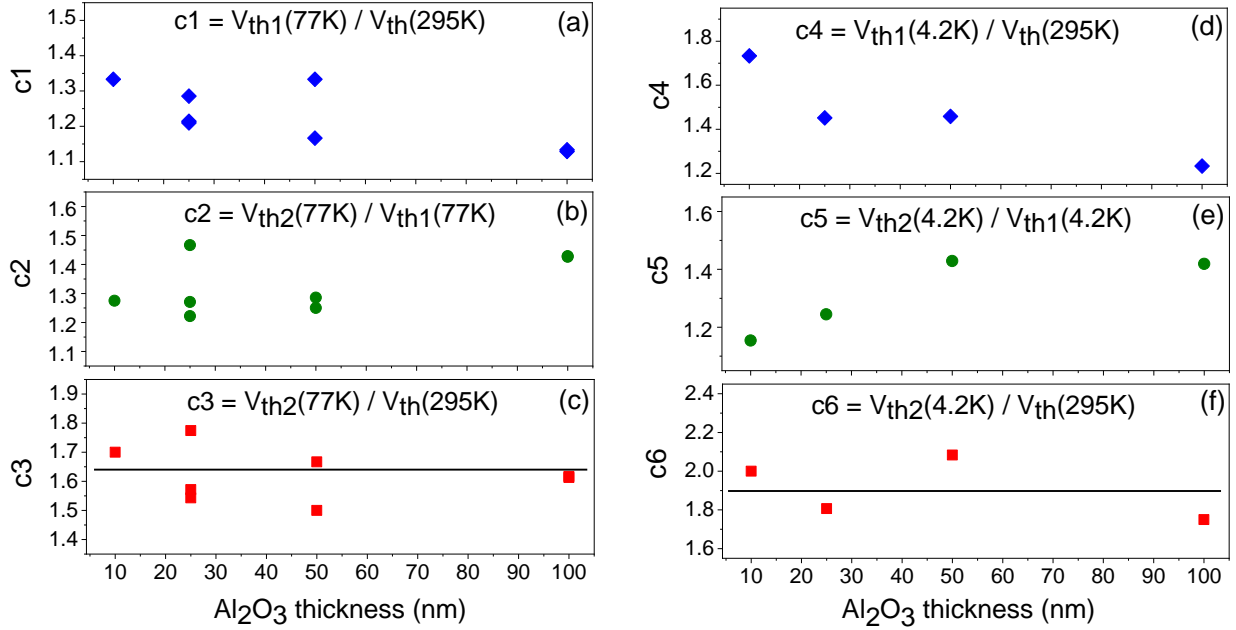

*Figure 4. Ratios of the threshold voltages (c1-c6) measured at cryogenic temperatures. Here $V_{th}(295K)$ represents the thresholds measured at the room temperature. Also, $V_{th1}(77K)$ and $V_{th1}(4.2K)$ represent, correspondingly, the thresholds measured at 77 K and 4.2 K in the first run of the V-I curve cycle, when the dielectric is not charged yet. Finally, $V_{th2}(77K)$ and $V_{th2}(4.2K)$ represent, correspondingly, the thresholds measured at 77 K and 4.2 K in the second and the subsequent runs of the V-I curve measurements, when the dielectric is charged. The horizontal line in (c) and (f) are shown to emphasize the approximate independence of the corresponding threshold ration to the film thickness. The ratios are defined as $c1=V_{th1}(77K)/V_{th}(295K)$, $c2=V_{th2}(77K)/V_{th1}(77K)$, $c3=V_{th2}(77K)/V_{th}(295K)$, $c4=V_{th1}(4.2K)/V_{th}(295K)$, $c5=V_{th2}(4.2K)/V_{th1}(4.2K)$, $c6=V_{th2}(4.2K)/V_{th}(295K)$.*

The discussion above was based on the assumption that, at cryogenic temperatures, the charges become trapped in the dielectric layer and also remain trapped even if the voltage is reduced to zero. Below we will discuss in detail how to explicitly confirm the fact that charges get trapped in the dielectric and will show that the charging effect is strongly nonlinear: The amount of charge injected and trapped in the dielectric is much larger if the charging voltage is chosen to be sufficiently high, so that the field emission current is detected. In this case, the electrons tunnel into the traps that are located near the anode (the positive plate), since these traps have lower energies [35,36].

Another important finding is that the hopping conductivity channel is completely suppressed at cryogenic temperatures, down to the precision of our measurements. This is the case because thermal energy is not sufficient to allow the electrons to jump from one trap to another. Thus, the energy storage is more reliable at cryogenic temperatures. The experiments discussed below will show that the energy remains stored in the dielectric even if the capacitor plates are short-circuited, which is an unexpected but important finding.

Below we will discuss how to probe dielectric-trapped charges directly, namely by warming up the sample and by simultaneously measuring the current generated by the electrons escaping from their traps. We term this phenomenon "thermal spectroscopy" of the trapped charges. It provides an estimate of the trapping energy.

The experimental circuit for performing the thermal spectroscopy is illustrated in Fig.1a, insert. The sample was charged at $V=5$ V for 18 h at 77 K. After the charging, the capacitor was discharged through a series standard resistor $R_{st1}=1$ G$\Omega$ for 5 min. The discharge then proceeded further, just to make sure that any charges that could move freely were fully discharged. To this goal, the plates of the capacitor have been short-circuited. Namely, the plates were connected using a copper wire for ~50 min. After the complete discharge, the capacitor was again connected to a current-measuring circuit and the temperature was slowly increased. The current was measured while no external voltage was applied.

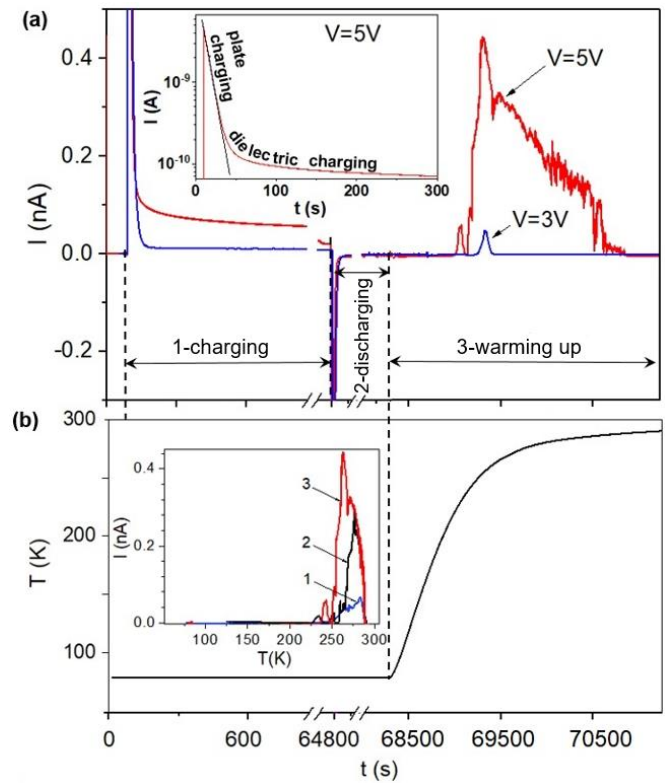

*Figure 5. (a) Current through a capacitor having a 10 nm dielectric film, as a function of time at different regimes: (1) charging (at 77 K), (2) discharge (at 77K), and (3) warming up to the room temperature. The sample was charged at 77 K at either V=5 V (red curve) or 3 V (blue curve) for 18 h. (a, insert) The charging current as a function of time. The black line represents the best exponential fit for the charging current of the plates of the capacitor, marked as "plate charging". At times t > 40 s the measured current is larger than the extrapolated exponential dependence. The observed excess current is due to the dielectric charging current (red curve marked as "dielectric charging"). (b) The temperature ramps up as a function of time. The horizontal time axis in (a) and (b) is the same. (b, insert) Current through the capacitor upon the warming up process as a function of a temperature. The sample was charged at 77 K at V=5 V for either 10 h (curve #1) or 16 h (curve #2), or 18 h (curve #3, data from (a), red curve). It is clear that the*

*charges remain unmoved up to ~225 K, but above that temperature the electrons leave their traps and quickly move to the anode.*

The dielectric charging current is explicitly measured during the charging process, as was reported in previous publications [26,34]. A typical charging curve is shown in Fig.5a (insert). As bias voltage is applied through a series resistor, the current is initially strong, but it drops exponentially with time. At some point, a "tail" is observed (approximately at $t < 40$ s), which indicates the presence of some excess current. This excess current is the signature of the dielectric charging.

In Fig.5 we present the current measured in the entire experimental sequence. The sequence of events includes the initial charging and then discharging of the capacitor performed at temperature $T$=77 K. During this voltage cycling, the electrons are able to enter the dielectric. But since the temperature is low, they become trapped inside the dielectric and form the Coulomb barrier as discussed above. Then, in the final stage of the test, the capacitor is warmed up, at zero applied voltage, and we are able to measure an additional current, $I(t)$, providing a direct probing tool for the electrons escaping from their trapping centers in the dielectric layer and exiting the device. A current increase was detected only when the temperature reached ~225 K, indicating that the trapped electrons are well trapped in the alumina below that temperature. As the temperature is further increased, the current goes to zero as all available electrons leave the dielectric layer.

Since $I=dQ/dt$, the integral of the current upon the warming up process equals the total charge stored in the dielectric. Thus, we obtain the total trapped charge: $Q$=4.6·10$^{-8}$ C, $Q$=1.4·10$^{-7}$ C, and $Q$=2.9·10$^{-7}$ C for the charging times of 10 hours, 16 hours, and 18 hours correspondingly (see the curves #1, #2 and #3 correspondingly, in Fig.5b(insert)). These results indicate that the amount of charge accumulated in the dielectric increases with the charging time.

The number of electrons released by the capacitor should be compared to the number of traps estimated in the above discussion of the hopping conductivity. The number of the released electrons can be evaluated as $N=Q/e$. The corresponding concentration of the trapped electrons is $n_e=N/V_{oxide}$, where $V_{oxide}=At_{oxide}=10^{-14}$ m$^3$ is the volume of the dielectric. Thus, the estimated concentration of the trapped electrons is $n_e=N/V_{oxide}\approx 3\cdot 10^{25}$ m$^{-3}$ or $\approx 8\cdot 10^{25}$ m$^{-3}$, or $\approx 1.8\cdot 10^{26}$ m$^{-3}$ (for the curve #1, curve #2 and curve #3 in Insert, Fig.5b, correspondingly). The trap concentration found from the above analysis of the hopping conductivity is very similar, namely $n_t\sim 10^{26}$ m$^{-3}$. Thus, we obtain additional evidence that at cryogenic temperatures the electrons permanently saturate the traps, namely the same traps which enable hopping conductivity at room temperature. The ability to store a large amount of charge in the insulator is based on the absence of hopping conductivity. This is in agreement with our finding that at cryogenic temperatures the stored charge is a few times larger than the charge stored on the capacitor plates and, at the same time, the hopping conductivity leakage is undetectable.

We have also attempted to charge the dielectric using a lower bias voltage at which no field emission leakage current can be detected. The exact conditions were $T$=77 K and $V$=3 V and the result is shown in Fig.5a as the blue curve, marked "V=3 V", to indicate the applied voltage. In this case the curve is qualitatively different than the "V=5 V" curve. One difference is that there is no detectable leakage current during the charging phase. The other, key difference is that upon warming up the capacitor charged at "V=3V" shows a charge release of only 3 nC, which is almost 100 times less then what was measured when the capacitor was charged at *V*=5 V. Thus, the charging effect is strongly nonlinear and requires an electron injection, which is only possible if the applied voltage is high.

## IV. DISCUSSION

To explain the results, we present a simple model (Fig.6). First, an electric field is appleid by biasing the capacitor to a voltage so high, that a field emission effect occurs and the electrons tunnel into the dielectic layer. (The voltage of course has to be lower than the breakdown voltage.) The left Al plate in the diagram is negatively charged, and the electrons penetrate into the dielectric forming a pile near the opposite electrode, the anode. This is because the tunneling effect energy has to be conserved. Thus, at votlages just slightly higher than the threshold at which the field emission begins, the electrons will tunnel into the region with the lowest trap energy, which is the region adjacent to the anode (Fig.6a).

If the voltage is increased, then the width of the classically forbidden region in the dielectric layer gets thinner. So the electrons occupy traps in a wide segment of the dielectric on the anode side. Also, the field emission current is stronger at higher voltages, so a larger percentage of traps gets occupied and the total charge trapped in the dielectric becomes much larger than in the case considered in Fig.6b.

Next, the voltage is set to zero (Fig.6c). Due to the trapping random potential, most of the electrons remain trapped in the dielectric film, assuming the temperature is still low, so that hopping conductivity is negligible. Thus, this arrangement represents an ideal energy storage environment, the electrons remain trapped even if the plates of the capacitor are short-circuited. Also, the trapped electrons present a Coulomb barrier for the leakage current.

In the last stage of the experiment, the temperature is increased, so that most of the electrons can escape from their trapping sites. They flow to the nearest electrode which is the anode. This experiment represents a form of thermal spectroscopy, since at each temperature the escaping electrons come from the traps whose energy depth is comparable to $k_BT$. At such temperatures, where electrons can escape from their traps, hopping conductivity begins. Thus, the electrons become mobile and move to the electrodes.

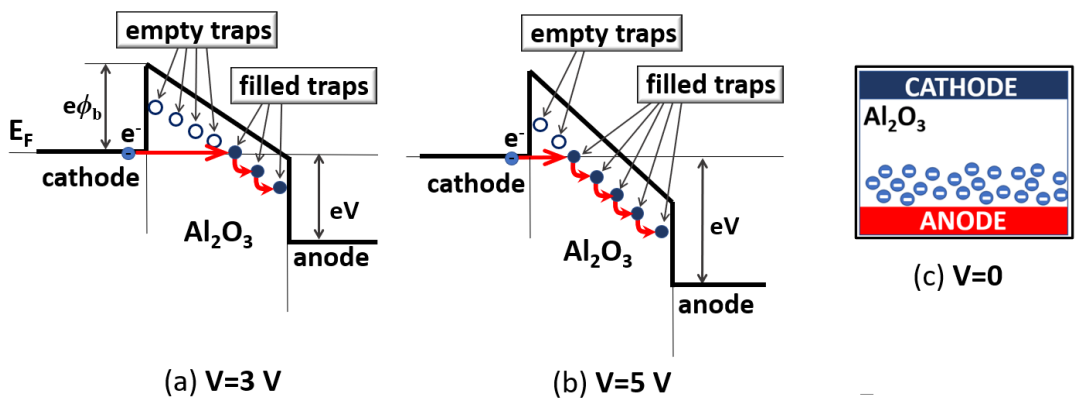

*Figure 6. A schematic illustrating the penetration of electrons (red color) in the dielectric. A negative potential is applied to the left electrode, driving the electrons inside the insulator by means of the field emission effect, which ensures that the electrons end up father away from the*

*cathode (the negative capacitor plate) since they can only tunnel into the region where their energy is equal or less than the Fermi energy in Al. In this case, the charge density is higher near the anode. Here $\phi_b$ is the metal-insulator energy barrier height for the tunneling electrons. (a) Weaker potential. Electrons can only tunnel to the states very close to the anode. (b) Stronger voltage bias. Electrons can tunnel in a wide segment of the dielectric closer to the anode. (c) The plates of the capacitor are shorted, so the voltage is zero. Many electrons remain trapped if the temperature is sufficiently low. If the temperature is increased, these trapped electrons would predominantly escape to the plate which is closer to them, i.e., to the anode.*

To summarize, our explanation to the observed changes of the V-I curves at cryogenic temperatures is that at high temperature the electrons injected into the dielectric can diffuse away by means of hopping conductivity. The injection is due to the field emission effect, which is discussed in the context of Fig.2a. The hopping conductivity effect is discussed in the context of Fig.2b, where a theoretical fit is given and matches the data. At low temperatures, the field emission is still present because it is a quantum mechanical tunneling effect, which is temperature independent. Yet the hopping conductivity, which is a thermally activated effect, freezes out (becomes very low). Thus, a qualitatively different phenomenon occurs at low temperatures, namely it becomes possible to inject electrons into the dielectric layer of the capacitor and build up a significant bulk charge there, which then acts as a Coulomb barrier for the leakage current.

## V. CONCLUSIONS

We have investigated the leakage current in capacitors with nanometer scale dielectric layers, ranging between 7 nm and 100 nm. Two main leakage mechanisms are identified. At higher electric fields, it is the field emission process which persists down to *T*=77 K and even *T*=4.2 K. The second leakage mechanism is the hopping conductivity, which is present at room temperature but becomes immeasurably low at cryogenic temperatures.

Our key finding is that repeated cycling of the voltage on capacitors at cryogenic temperatures changes the properties of their dielectric spacer, such that the leakage current is reduced. Our conclusion is that the electrons can penetrate the insulator layer when the voltage is such that the field emission current between the plates of the capacitor just begins. The electrons then can tunnel into the traps, which are predominantly located near the anode. The charge builds up in the dielectric and causes a Coulomb barrier of the leakage current. The field emission is still observed but its onset is shifted to significantly higher voltages. Such charging of the dielectric is stable only if the temperature is sufficiently low, so that the accumulated charge cannot diffuse away by means of hopping conductivity.

We have confirmed the trapping of the electrons in the dielectric film by means of thermal spectroscopy: if the sample is slowly warmed up, a significant current is detected, which flows in the *same* direction as the charging current. The fact that the direction of the thermally released current coincides with direction of the current observed during the initial charging confirms the fact that the electrons are present in the dielectric, as is illustrated in Fig.6. If some sort of polarization would be responsible, then it would generate a current in the opposite direction [37]. The number of the released electrons, upon the warming, matches the estimated number of traps.

Based on these results, we can envision two types of procedures which might improve the energy storage abilities and reduce the leakages in capacitors with nanoscale dielectric films. The first one is to operate these devices at cryogenic temperatures, where the induced Coulomb barrier is stable after a conditioning cycle. The second is to search or design insulators in which the energy of the electronic traps is higher, so that the hopping conductivity would be suppressed even at room temperature.

**ACKNOWLEDGMENTS**
This work was supported by the Air Force grant AF FA9453-18-1-0004.